# Strong circular dichroism in twisted single-ring hollow-core photonic crystal fiber

P. Roth[1], Y. Chen[1], M. C. Günendi[1], R. Beravat[1], N. N. Edavalath[1], M. H. Frosz[1], G. Ahmed[1], G. K. L. Wong[1,*], P. St.J. Russell[1,2]

1 Max Planck Institute for the Science of Light, Staudtstr. 2, 91058 Erlangen, Germany
2 Department of Physics, University of Erlangen-Nuremberg, Staudtstr. 2, 91058 Erlangen, Germany
*Corresponding author: gordon.wong@mpl.mpg.de

**We report a series of experimental, analytical and numerical studies demonstrating strong circular dichroism in helically twisted hollow-core single-ring photonic crystal fiber (SR-PCF), formed by spinning the preform during fiber drawing. In the SR-PCFs studied, the hollow core is surrounded by a single ring of non-touching capillaries. Coupling between these capillaries results in the formation of helical Bloch modes carrying orbital angular momentum. In the twisted fiber, strong circular birefringence appears in the ring, so that when a core mode with a certain circular polarization state (say LC) phase-matches to the ring, the other (RC) is strongly dephased. If in addition the orbital angular momentum is the same in core and ring, and the polarization states are non-orthogonal (e.g., slightly elliptical), the LC core mode will experience high loss while the RC mode is efficiently transmitted. The result is a single-circular-polarization SR-PCF that acts as a circular polarizer over a certain wavelength range. Such fibers have many potential applications, for example, for generating circularly polarized light in gas-filled SR-PCF and realizing polarizing elements in the deep and vacuum ultraviolet.** © 2018 Optical Society of America

## 1. INTRODUCTION

Solid-core photonic crystal fibers (PCFs) that support only one linear polarization state have been realized in various different ways, for example by raising the effective index of one of the core modes in a birefringent PCF above that of the highest index space-filling cladding mode [1,2], or by creating different cut-off wavelengths for orthogonal linearly polarized modes in an all-glass hybrid photonic bandgap PCF [3].

It has been known since the 1980s that spinning the preform during fiber drawing creates elliptical birefringence in a fiber with a linearly birefringent core [4], and perfect circular birefringence in a fiber with a non-circular *N*-fold rotationally symmetric core [5], as recently demonstrated in solid-core PCF [6]. Although it has been proposed that twisted linearly-birefringent solid-core PCF can be designed to support just one elliptically polarized mode [7], this has not yet to our knowledge been realized experimentally. In this paper we report a twisted single-ring (SR) hollow-core PCF that transmits one circular polarization state with high efficiency, while the other experiences high loss. To our knowledge this is the first ever report of an all-glass optical fiber that exhibits circular dichroism.

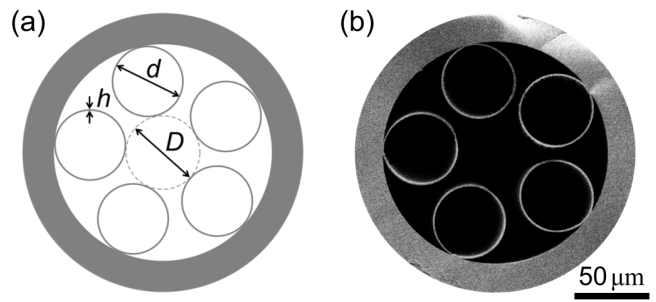

Fig. 1. (a) Schematic of the fiber cross-section. The hollow core with diameter *D* is surrounded by five evenly spaced capillaries of wall thickness *h* and inner diameter *d*. (b) Scanning electron micrograph (SEM) of the twisted SR-PCF used in the experiments. The helical twist period is 1.19 cm, *d* = 45 µm, *D* = 50 µm and *h* = 1 µm.

SR-PCF is an emerging class of hollow-core fiber consisting of a ring of anti-resonant elements (e.g., capillaries of diameter *d*) surrounding a central hollow core (diameter *D*) [8-11]. It has many applications, for example, in high power delivery of laser light [12], ultrafast gas-based nonlinear interactions [13] and chemical sensing [14]. SR-PCF guides by anti-resonant reflection, which relies on the core mode being dephased from the modes of the surrounding capillaries so that it cannot strongly couple to them. When in addition the geometrical "shape" parameter *d/D* = 0.682, higher-order core modes are strongly suppressed over a broad bandwidth while low-loss guidance of the fundamental LP$_{01}$-like mode is maintained [15]. This condition can be tuned by spinning the preform during fiber drawing to form a multi-helix of hollow channels [16] or by bending the fiber to a suitably chosen radius of curvature [17].

Twisting has the additional benefit that, above a certain minimum twist-rate, the left- and right-circularly polarized (LCP and RCP) constituents of a linearly polarized mode (for example, the HE$_{11}$–like core mode) become non-degenerate, with the result that the fiber becomes optically active and polarization-preserving for circularly polarized light [18]. This is a key ingredient in designing a unique hollow-core SR-PCF that exhibits strong circular dichroism, i.e., over a certain bandwidth it transmits only LCP or RCP light, depending on the handedness of the chirality.

## 2. PHYSICAL MECHANISM

In this section we discuss the necessary and sufficient conditions for coupling to occur between core and ring modes, in particular how one circular polarization state can couple, experiencing high loss, while the orthogonal state does not couple and so is efficiently transmitted.

### A. Cylindrical versus Cartesian coordinates

In a conventional fiber with a circular core, the HE$_{11}$ mode is formed from the superposition of a pair of circularly polarized modes whose orbital angular momentum (OAM) $\ell$ and spin $s$ are $(\ell, s) = (\pm 1, \mp 1)$. These modes are defined in cylindrical coordinates (their natural frame of reference, $(\rho, \phi, z)$ forming a right-handed set) with transverse electric fields:

$$\mathbf{e}_{\text{cyl}} = (e_\rho, e_\phi)^T = \frac{e_0}{\sqrt{2}} (1, is)^T e^{i\ell\phi} \qquad (1)$$

where $\phi$ is the azimuthal angle and $\rho$ the radius. The field components in Cartesian coordinates are then:

$$\mathbf{e}_{\text{crt}} = \begin{pmatrix} \cos\phi & \sin\phi \\ -\sin\phi & \cos\phi \end{pmatrix} \cdot \mathbf{e}_{\text{cyl}} = \frac{e_0}{\sqrt{2}} (1, is)^T e^{i(\ell+s)\phi}, \qquad (2)$$

which shows that, for a given mode, the OAM orders in the cylindrical and Cartesian frames differ by $s$.

When calculated in a Cartesian frame, the vector fields of an eigenmode in a twisted SR-PCF are aperiodic with azimuthal angle. When evaluated in the cylindrical frame, however, they turn out to repeat exactly in each azimuthal period (see numerical modeling in Section 2E), with a phase advance per period that equals $2\pi\ell/N$. In other words, the fields form perfect helical Bloch modes when expressed in cylindrical coordinates.

### B. Untwisted SR-PCF

The core modes in the untwisted SR-PCF are very similar to the pair of circularly-polarized modes mentioned in the previous section, i.e., their OAM and spin are $(\ell, s) = (\pm 1, \mp 1)$. The modes of the ring of coupled capillaries are rather more complicated. Their OAM-carrying properties may be understood by using vector coupled-mode theory in a cylindrical coordinate system to treat a ring of $N$ coupled cores, and then applying Bloch's theorem (see Appendix). For the special case when the inter-core coupling $\kappa$ is independent of polarization state and the cores have zero birefringence, the dispersion relation takes the simple form (useful for illustrative purposes):

$$\gamma = 2\kappa \cos(2(\ell + s)\pi / N) \qquad (3)$$

where $\gamma$ is the correction to the axial propagation constant $\beta_0$ of the mode in an isolated core (see Appendix, section A). Two modes with the same OAM order $\ell$ but opposite spins thus differ in refractive index by:

$$B_C(\ell) = (2\kappa\lambda/\pi)\sin(2\pi/N)\sin(2\pi\ell/N). \qquad (4)$$

Within the first Brillouin zone $-N/2 < \ell_0 < N/2$, the need to satisfy an azimuthal resonance restricts $\ell$ to integer values. The azimuthal periodicity also creates Bloch harmonics with OAM orders $\ell_m = \ell_0 + mN$ where $m$ is an integer. In the untwisted fiber, all these orders share the same axial propagation constant and group velocity for a given value of spin.

Note that if the inter-core coupling rate is polarization-dependent, or the cores are birefringent, the polarization state of the eigenmodes is not perfectly circular (see Appendix section A).

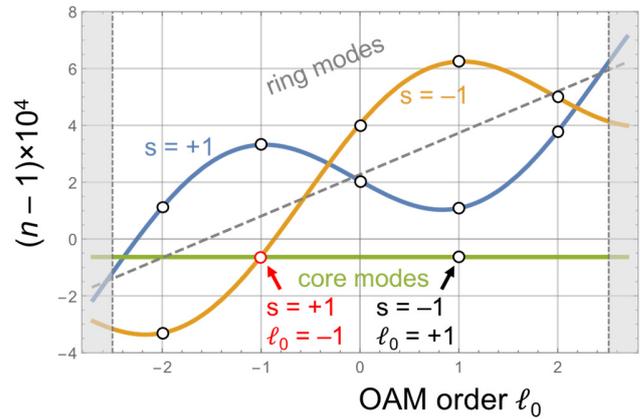

Fig. 2: Illustrative plot of modal refractive index versus OAM order within the first Brillouin zone of a twisted SR-PCF, showing core-to-ring phase-matching (note that the circular birefringence of the core modes is very small: $\sim 10^{-7}$). The values of OAM allowed in the ring are marked with circles. The $s = +1$, $\ell_0 = -1$ core mode phase-matches to the $s = -1$, $\ell_0 = -1$ ring mode (red arrow), while the $s = -1$, $\ell_0 = +1$ core mode is not phase-matched to any ring mode.

### C. Twisted SR-PCF

When the fiber is helically twisted the two core modes become non-degenerate [19], i.e., the LCP and RCP modes develop very slightly different propagation constants. Whereas this is a relatively small effect in the core ($B_C \sim 10^{-7}$), it is much stronger in the ring of coupled capillaries. The result is that only one of the circularly-polarized core modes is able to phase-match to a ring-mode and thus exhibit high loss, while the other core mode remains unaffected.

The twisted fiber can be analyzed by applying a suitable coordinate transformation as explained in [18], yielding a closed form expression for the modal refractive index:

$$n(\ell_m) = n_0\left(1 + \frac{\alpha^2\rho^2}{2}\right) + \frac{\ell_m \alpha \lambda}{2\pi}$$
$$+ 2n_\kappa \cos\left(\frac{2\pi}{N}(\ell_m + s - 2\pi n_0 \alpha \rho^2 / \lambda)\right) \quad (5)$$

where $\lambda$ is the vacuum wavelength, $n_0 = \beta_0\lambda/2\pi$ is the index of the mode in an isolated capillary, $n_\kappa = \kappa\lambda/2\pi$, $\alpha$ is the twist rate in rad/m, $\rho$ is the radial distance of the capillary centers from the fiber axis and the approximation $\alpha^2\rho^2 \ll 1$ has been used ($|\alpha\rho| \approx 0.001$ in the experiments reported here). The first term in Eq. (5) takes account of the fact that the average effective refractive index along the fiber axis is increased by a factor $(1 + \alpha^2\rho^2)^{1/2}$ because of the helical path taken by the light.

To illustrate how the dispersion relation in Eq. (5) can give rise to circular dichroism, in Fig. 2 we plot modal refractive indices versus OAM order within the first Brillouin zone for a SR-PCF with $N = 5$, $\alpha\rho = 0.0262$, $n_0 = 0.999884$, $\rho/\lambda = 28.5$, $n_\kappa = 0.000136$ and the index of both core modes is 0.999937 (these parameter values correspond approximately to those in the experiment). The plot shows that it is possible to find conditions when one of the circularly polarized core modes is phase-matched to the ring, while the orthogonal one is not.

### D. Phase-matching, selection rules and orthogonality

Three conditions must be satisfied for coupling to occur between core and ring:

1. The modal refractive indices must match (for phase-matched coupling);
2. The OAM orders $\ell_m$ must be the same in core and ring (for non-zero overlap integral around the azimuth);
3. The polarization states of the core and ring modes must be non-orthogonal (at least to some extent).

The third condition is not satisfied in the simplified model used for Fig. 2, when core and ring modes are circularly polarized and orthogonal. Numerical modeling (next sub-section) reveals, however, that the fields have complex spatially-varying polarization states in the twisted SR-PCF, allowing core and ring modes to couple.

### E. Polarization state as function of azimuthal angle

Solving Maxwell's equations in a helicoidal coordinate frame [20,21] enables us to calculate all four Stokes parameters (referred to the cylindrical frame) as a function of angle around the azimuth in a geometrically perfect SR-PCF. The results are plotted in Fig. 3 for the lower loss core+ring mode labelled (ii) in Fig. 4, close to the phase-matching wavelength of 1593 nm. It is remarkable to observe that all the Stokes parameters evolve periodically around the azimuth, with $N$ periods per round-trip. This is true at every value of radius, as illustrated for $\rho/D = 0.25$, 0.50 and 1.05 in the figure. The fields thus form a perfect helical Bloch mode with a phase advance between adjacent periods of $2\pi\ell_m/N$ for the $m$-th Bloch harmonic. Note in particular that the polarization state is far from circular (except close to the core center), creating the conditions necessary to fulfill the third condition in the previous subsection.

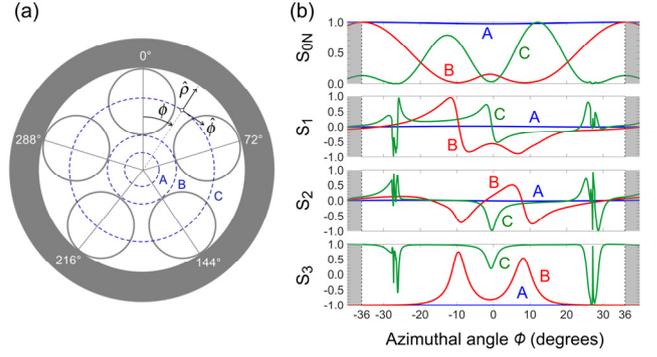

Fig. 3: (a) Sketch showing the circular paths along which the Stokes parameters are calculated relative to the local $(\hat{\rho}, \hat{\phi})$ axes at three different values of radius: $\rho/D = 0.25$ (A), 0.5 (B) and 1.05 (C). (b) Stokes parameters for mode (ii) in Fig. 4, plotted against azimuthal angle $\phi$ over one period. The curves are identical in each period of the ring, as expected of a helical Bloch wave. $S_{0N}$ is the modulus squared of the electric field, normalized to its maximum value $S_{0M}$ along a given path ($S_{0M} = 1$ on axis, 0.5043 for A, 0.0350 for B and 0.0244 for C). When $\mathbf{S} = (S_1, S_2, S_3) = (-1,0,0)$ the fields are linearly polarized in the radial direction; when $\mathbf{S} = (1,0,0)$, linearly polarized in the azimuthal direction; when $\mathbf{S} = (0,1,0)$, linearly polarized at $\phi = \pi/4$ to the radial direction; and when $\mathbf{S} = (0,-1,0)$, linearly polarized at $\phi = -\pi/4$ to the radial direction. The light is LCP when $\mathbf{S} = (0,0,1)$ and RCP when $\mathbf{S} = (0,0,-1)$.

## 3. EXPERIMENTAL AND NUMERICAL RESULTS

The fiber was fabricated using a modified version of the two-step stack-and-draw technique. It consisted of a central hollow core with inner diameter $D = 50$ μm surrounded by five evenly spaced non-touching capillaries with wall thickness $h = 1.0$ μm and inner diameter $d = 45$ μm (see scanning electron micrograph in Fig. 1(b)). The shape parameter $d/D$ was thus 0.9. A permanent twist (twist rate 529 rad/m) was imposed by spinning the fused-silica preform around its axis while drawing it down to fiber [16]. In the resulting SR-PCF the capillaries follow a helical path around the core.

Figure 4 shows the calculated effective index and loss in the laboratory frame of the RCP core+ring eigenmodes near phase-matching point, when all three conditions in Section 2D are fulfilled. The modelled structure is perfectly five-fold symmetric, so as to ensure perfect azimuthal periodicity. The normalized Poynting vector distributions of the two core+ring eigenmodes are also shown in Fig. 4. It is interesting that an anti-crossing in loss, not refractive index, is seen. This occurs because the coupling rate $\kappa_{CR}$ between core and ring is less than $2\alpha$, where $\alpha$ is the amplitude decay rate in the ring (see Appendix, section B).

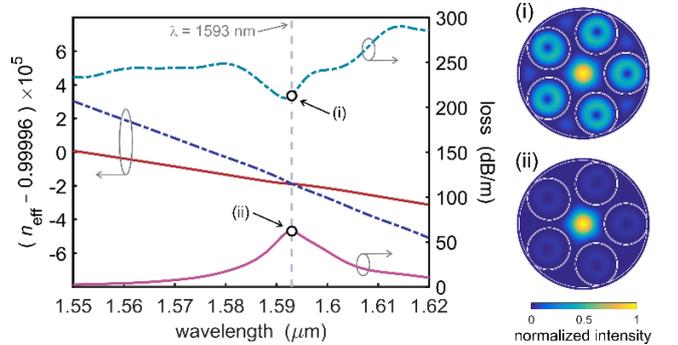

Fig. 4. Left-hand side: Modal indices and loss in the laboratory frame as a function of wavelength for the two core+ring eigenmodes of the twisted fiber,

calculated by finite element modeling. As one moves away from the phase-match wavelength (1593 nm), these modes evolve into core (solid curves) and lossy ring (dotted curves) modes. The modeled structure is ideal (see text) and the twist rate 529 rad/m. In this case only the ring and the RCP core modes satisfy all three conditions described in section 2D, resulting in resonant coupling. Right-hand side: Poynting vector distributions for the (i) high and (ii) low loss eigenmodes of core+ring at the phase-match wavelength (note that the loss rate is higher than the coupling rate, resulting in an anti-crossing in loss, not index – see Appendix, section B).

The cut-back method was used to measure the attenuation spectra of LCP and RCP core modes, using a linearly polarized supercontinuum laser as light source. The input polarization state was controlled with a quarter-wave plate, and the launch conditions were kept constant throughout the experiment. The twisted single-ring HC-PCF was kept straight and its optical output was coupled into a single-mode fiber, connected to an optical spectrum analyzer. The experiment was carried out with fiber lengths from 2.5 m to 1.2 m and the results are shown in Fig. 5(a), revealing a strong circularly dichroic spectral band. As expected from the theory, near the phase-match wavelength (1593 nm) only the RCP core mode is resonantly coupled to the lossy ring mode in the spiraling capillaries, while the LCP core mode is unaffected. Far from this wavelength, both polarizations exhibit similar loss. A maximum attenuation of 9.7 dB/m at 1590 nm was achieved for RCP light, while the loss for LCP light was 1.4 dB/m. The overall spectral dependence of circular dichroism shows good agreement with the calculations. With these parameters, a 3 m length of the twisted SR-PCF will transmit 0.1% of RCP light compared to 38% of LCP light.

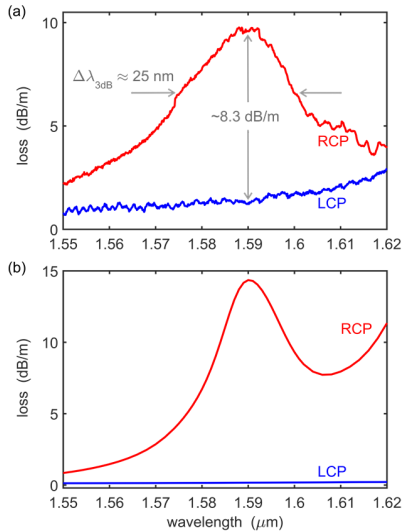

Fig. 5. (a) Measured and (b) numerically modeled modal loss of the RCP and LCP core modes as a function of wavelength for the twisted SR-PCF. A scanning electron micrograph of the actual fiber structure was used in the modeling. As a result the calculated RCP modal loss is about 3 times lower than in the ideal structure (Fig. 4), although it agrees reasonably well with the measured value.

The 3-dB single-circular-polarization bandwidth is 25 nm, within which the fiber transmits only the LCP core mode after some length of propagation (Fig. 5). We attribute the larger measured circular dichroism bandwidth (compared to the calculated value) to small deviations from the ideal structure, not taken account of in the modeling, which cause inhomogeneous broadening in the experiment.

At wavelengths far from the dichroism band the circular birefringence $B_C$ of the core modes was measured by launching linearly polarized light and monitoring the sign and magnitude of the polarization rotation; the resulting circular birefringence was of order $10^{-7}$.

An interferometric setup was used to measure the spatial phase and polarization distributions of the modes. Light from a 1550 nm laser was split and diverted into reference and measurement arms using two computer-generated holograms placed at different positions on a phase-only spatial light modulator. The reference beam, with a planar wavefront, was combined with the "image" light emerging from the fiber so as to generate an interference pattern on an imaging sensor. The relative phase between reference and "image" was varied from 0 to $2\pi$ in equidistant steps by phase-modulating the reference arm hologram. At each pixel we recorded the relative phase at which maximum constructive interference was observed, permitting reconstruction of the phase distribution across the entire mode. By adding a polarizing beam-splitter at the output, the relative phase distributions of horizontal and vertical electric field components could be measured simultaneously using two cameras. The phase difference between these two components allows the distribution of polarization state to be mapped across the optical near-field pattern at the fiber end-face. The measured and calculated results for the high-loss RCP core mode at a wavelength of 1550 nm are plotted in Fig. 6, showing reasonably good agreement.

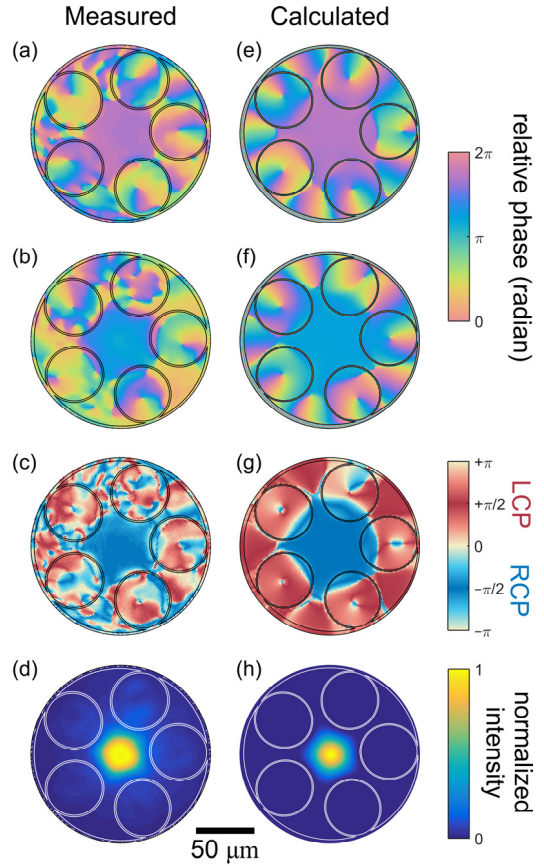

Fig. 6. Measured phase distribution of the (a) horizontal and (b) vertical electric field components of the high-loss RCP core mode at a wavelength of 1550 nm (the LCP core mode has low loss). The fiber length was 1.2 m. (c)

Measured polarization distribution derived from the phase difference of (a) and (b). (d) Measured normalized intensity distribution. (e)-(h) Corresponding numerically calculated distributions, based on a scanning electron micrograph of the actual fiber structure.

An attractive feature of twisted SR-PCF is that the field distribution of the fundamental mode in the core retains a high degree of azimuthal symmetry as well as a large mode field area. Figure 6(d) shows the measured near-field intensity distribution of the core mode at 1550 nm, when the mode field diameter is ~35 μm. Light is well confined to the core and emerges in a Gaussian-like mode. Higher-order core modes have very high loss, rendering the twisted SR-PCF effectively single-mode as well as circularly dichroic.

## CONCLUSIONS

SR-PCFs can be made circularly dichroic by twisting during fiber drawing. Depending on the sign of twist, light in a given circular polarization states couples strongly to leaky modes in the capillary ring, while orthogonally polarized light is efficiently transmitted. A loss discrimination of 8.3 dB/m between RCP and LCP core modes at 1590 nm was achieved (1.4 dB/m for RCP light and 9.7 dB/m for LCP light) in the first experimental demonstration of this novel effect. The central wavelength of the circular dichroism band can be tuned over a broad range by varying $d/D$ and the twist rate. Such fibers may be useful in many contexts, for example in generating pure circularly polarized light in the deep and vacuum UV, photochemistry of chiral isomeric molecules, elimination of polarization mode dispersion and polarization instability and realizing polarizing elements in the deep and vacuum ultraviolet. The large mode area could also be attractive for high power circularly polarized fiber lasers and amplifiers.

## APPENDIX

### A. Vector coupled-mode theory

Here we show how the vector dispersion relation of the modes in a ring of $N$ cores can be derived in the untwisted case. Figure A1 sketches the field components in cylindrical coordinates for the $n$th, $(n-1)$th and $(n+1)$th cores in the ring. Resolving the fields in the $n$th and $(n+1)$th cores along the $(\rho, +\pi/N)$ axes, and those in the $n$th and $(n-1)$th cores along the $(\rho, -\pi/N)$ axes, results in the following equation for nearest-neighbor coupling:

$$\dot{\mathbf{e}}_n = i[\mathbf{D}]\mathbf{e}_n + i\left([\mathbf{R}]^{-1}[\mathbf{K}][\mathbf{R}]^{-1}\mathbf{e}_{n+1} + [\mathbf{R}][\mathbf{K}][\mathbf{R}]\mathbf{e}_{n-1}\right) \quad (A1)$$

where $\mathbf{e}_n = (e_n^\rho \hat{\rho} + e_n^\phi \hat{\phi})$ and the rotation and coupling matrices are:

$$[\mathbf{R}] = \begin{pmatrix} \cos\phi_N & \sin\phi_N \\ -\sin\phi_N & \cos\phi_N \end{pmatrix}, \quad [\mathbf{K}] = \begin{pmatrix} \kappa_1 & 0 \\ 0 & \kappa_2 \end{pmatrix} \quad (A2)$$

with $\phi_N = \pi/N$. Birefringence in the individual ring cores is modeled by:

$$[\mathbf{D}] = \begin{pmatrix} 2\pi B/\lambda & 0 \\ 0 & 0 \end{pmatrix} \quad (A3)$$

where $B$ the linear birefringence. Applying Bloch's theorem, we obtain the recurrence relation:

$$i\gamma\mathbf{e}_n = i\left([\mathbf{D}] + [\mathbf{R}]^{-1}[\mathbf{K}][\mathbf{R}]^{-1}e^{-i\phi_B} + [\mathbf{R}][\mathbf{K}][\mathbf{R}]e^{i\phi_B}\right)\mathbf{e}_n \quad (A4)$$

where $\phi_B$ is the Bloch phase between adjacent cores. Requiring the round-trip phase to be a multiple of $2\pi$ restricts $\phi_B$ to discrete values given by $2\pi\ell/N$. Eq. (A4) leads to the dispersion relation in Eq. (1) for $B = 0$ and $\kappa_1 = \kappa_2 = \kappa$, under which circumstances the eigenmodes are perfectly circularly polarized. In the SR-PCF it is highly likely that $\kappa_1 \neq \kappa_2$ and $|B| > 0$. As shown in Section 2E, this results in Stokes parameters that, when referred to a cylindrical frame, vary periodically with azimuthal angle, repeating $N$ times around the azimuth.

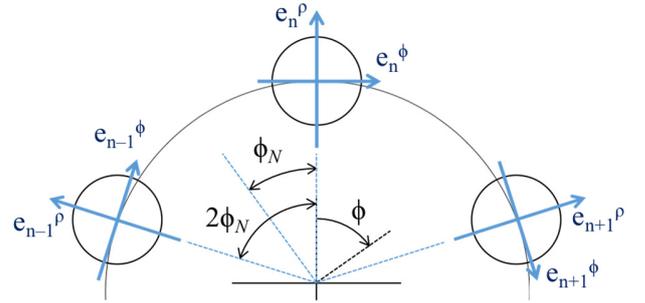

Fig. A1: Illustrating the field definitions in vector coupled-mode theory, used to derive the dispersion relation for azimuthal Bloch waves. The angular width of an azimuthal period is $\phi_N = \pi/N$.

### B. Coupled mode theory in presence of loss

The absence of an anti-crossing in refractive index for coupling between core and ring (Fig. 4) may be simply explained using coupled mode theory, assuming a lossless core mode and a high loss ring mode. The equations take the form:

$$\begin{aligned}\dot{a}_C &= i\gamma a_C = i\kappa_{CR} a_R \\ \dot{a}_R &= i\gamma a_R = i\kappa_{CR} a_C + (-\alpha_R + i\vartheta)a_R\end{aligned} \quad (A5)$$

where $a_R$ and $a_C$ are the modal amplitudes in ring and core, $\gamma$ is the coupling-related correction to the propagation constant, $\kappa_{CR}$ is the coupling constant, $\vartheta$ the wavelength-dependent dephasing and $\alpha_R$ the loss of the ring mode. At exact phase-matching ($\vartheta = 0$) the system has eigenvalues and vectors:

$$i\gamma = \alpha/2 \pm \sqrt{\alpha^2/4 - \kappa^2},$$
$$\left(a_C, a_R\right)_{\text{eigen}} = \left(-\frac{\alpha}{2\kappa} \pm \sqrt{\left(\frac{\alpha}{2\kappa}\right)^2 - 1}, -i\right). \quad (A6)$$

If $\alpha > 2\kappa$ the eigenvalues are entirely real, i.e., there is an anti-crossing in loss but not index, the high loss eigenmode (+ sign) being concentrated in the ring while the low loss mode (– sign) is concentrated in the core. This corresponds to the situation in Fig. 4.

**Funding.** Max Planck Society (MPG).